\documentclass[twocolumn]{book}
\usepackage[dvips]{graphicx,color}
\usepackage{makeidx,universe}

\makeauthorindex

\BookTitle{Proceedings of the XXIX PHYSICS IN COLLISION}

\CopyRight{\copyright 2009 by Universal Academy Press, Inc.}

\begin{document} 

\pagenumbering{arabic}

\chapter{%
{\LARGE \sf
Exotic Hadrons} \\
{\normalsize \bf 
Chang-Zheng Yuan$^{1}$
on behalf of Belle Collaboration} \\
{\small \it \vspace{-.5\baselineskip}
(1) Institute of High Energy Physics, Chinese Academy of Scineces,
Beijing 100049, China} }




  \baselineskip=10pt 
  \parindent=10pt    

\section*{Abstract} 

In this talk, we review the most recent progress in the searching
for the exotic hadrons, including hybrids, multi-quark states,
molecules and so on. We only focus on the studies with a
charmonium and one or more light mesons in the final states. This
covers the $X(3872)$, the $XYZ$ states at around 3.940~GeV, the
$Y(4140)$ and $X(4350)$ in two-photon collisions, the $Y$ states
from $ISR$ processes, and the charged $Z$ states.

\section{Introduction} 

In the quark model, mesons are composed from one quark and one
anti-quark, while baryons are composed from three quarks. Although
no solid calculation shows hadronic states with other
configurations must exist in QCD, people believe hadrons with no
quark (glueball), with excited gluon (hybrid), or with more than
three quarks (multi-quark state) exist. Since a proton and a
neutron can be bounded to form a deuteron, it is also believed
other mesons can also be bounded to produce molecules.

It is a long history of searching for all these kinds of states,
however, no solid conclusion was reached until now on the
existence of any one of them, except deuteron.

As the $B$-factories accumulate more and more data, lots of new
states have been observed in the final states with a charmonium
and some light hadrons. All these states populate in the
charmonium mass region. They could be candidates for usual
charmonium states, however, there are also lots of strange
properties shown from these states, these make them more like
exotic states rather than conventional states.

In this talk, we show the most recent results on the study of the
$X(3872)$, the $XYZ$ states at around 3.940~GeV, the $Y(4140)$ and
$X(4350)$, the $Y$ states from $ISR$ processes, and the charged
$Z$ states. The $X(3915)$ and $X(4350)$ found in two-photon
processes, and the $Y(4140)$ found in $B$ decays are new
observations.

\section{The $X(3872)$}

The $X(3872)$ was discovered by Belle in 2003~\cite{belle_x3872}
as a narrow peak in the $\pi^+\pi^-J/\psi$ invariant mass
distribution from $B\to K \pi^+\pi^-J/\psi$ decays. This discovery
mode was remeasured with more statistics both at Belle and at
BaBar. Belle reported a new result for the mass of the $X(3872)$
as $M^{Belle}_{X(3872)} = 3871.46\pm 0.37\pm
0.07$~MeV~\cite{belle_x3872_mass}; and BaBar~\cite{babar_deltam}
measured $M^{BaBar}_{X(3872)} = 3871.30\pm 0.60\pm 0.10$~MeV. the
most precise measurement of the mass was reported by CDF using the
same decay channel: $M^{CDF}_{X(3872)} = 3871.61\pm 0.16\pm
0.19$~MeV~\cite{CDF_x3872_mass}. A new world average that includes
these new measurements plus other results that use the $\pi^+\pi^-
J/\psi$ decay mode is $M^{\rm avg}_{X(3872)} = 3871.46\pm
0.19$~MeV, which is very close to the $D^{*0}\bar{D^0}$ mass
threshold: $m_{D^{*0}}+m_{D^0} = 3871.81 \pm 0.36$~MeV~\cite{PDG}.
This suggests a binding energy of $-0.35\pm 0.41$~MeV if $X(3872)$
is interpreted as a $D^{*0}\bar{D^0}$ molecule. Both Belle and
BaBar measured the mass difference for the $X(3872)$ produced in
neutral and charged $B$-meson decays. They both find mass
differences that are consistent with zero: $M^{B^+}_{X} -
M^{B^0}_{X} = 0.2\pm 0.9\pm 0.3$~MeV for Belle and $2.7\pm 1.6\pm
0.4$~MeV for BaBar. The CDF group tried to fit the $X(3872)\to
\pi^+\pi^- J/\psi$ peak with two states, they ruled out a mass
difference of less than 3.6~MeV (95\% C.L.) for two $X$ states
with equal production rate~\cite{CDF_x3872_mass}. These results do
not support the interpretation of the $X(3872)$ as a tightly bound
diquark-diantiquark system~\cite{maiani_1,ebert}, which expects
two nearby states with mass difference of $8\pm 3$~MeV.

With a data sample containing 447M $B\bar{B}$ events, Belle
observed a near-threshold $D^0\bar{D^0}\pi^0$ mass enhancement in
$B\to K D^0\bar{D^0}\pi^0$ decays that, when interpreted as
$X(3872)\to D^0\bar{D^0}\pi^0$, gave an $X(3872)$ mass of
$3875.4\pm 0.7^{+1.2}_{-2.0}$~MeV~\cite{belle_x3872_ddpi}. BaBar
studied $B\to KD^{*0}\bar{D^0}$ with a sample of 383M $B\bar{B}$
pairs and found a similar near-threshold enhancement that, if
considered to be due to the $X(3872)\to D^{*0}\bar{D^0}$, gave a
mass of $3875.1^{+0.7}_{-0.5}\pm
0.5$~MeV~\cite{babar_x3872_ddstr}. This state has been considered
to be a state different from the $X(3872)$ in literatures.
However, a subsequent Belle study of $B\to KD^{*0}\bar{D^0}$ based
on 657M $B\bar{B}$ pairs finds a mass for the near threshold peak
of $3872.9^{+0.6+0.4}_{-0.4-0.5}$~MeV~\cite{belle_ddstar} by
fitting the peak with a phase-space modulated Breit-Wigner (BW)
function, much closer to the value determined from the $\pi^+\pi^-
J/\psi$ decay channel.

The quantum number of the $X(3872)$ was found to be $1^{++}$
preferred. A study of the $\pi^+\pi^-$ mass distribution and the
observation of its $\gamma J/\psi$ decays~\cite{belle_gpsi}
indicate the $C$-parity of the $X(3872)$ is even, and the angular
correlations among the $\pi^+\pi^-J/\psi$ final state particles
constrains the $J^{PC}$ for the $X(3872)$ to be $1^{++}$ or
$2^{-+}$, with $1^{++}$ preferred~\cite{CDF_jpc}. Subsequently,
the $2^{-+}$ assignment has been further disfavored by BaBar's
report of $>3\sigma$ significance signals for $X(3872)$ decays to
both $\gamma J/\psi$ and $\gamma\psi(2S)$~\cite{babar_gammajpsi}.
The radiative transition of a $2^{-+}$ state to the $J/\psi$ or
$\psi(2S)$ would have to proceed via a higher order multipole term
and be highly suppressed. For these reasons, the most likely
$J^{PC}$ of the $X(3872)$ is $1^{++}$. The branching fraction of
$X(3872)\to \gamma \psi(2S)$ is found to be larger than that of
$X(3872)\to \gamma J/\psi$~\cite{babar_gammajpsi}, this is in
contradiction with the molecule interpretation of the $X(3872)$
state~\cite{swanson}.

Belle did a study of $X(3872)$ production in association with a
$K\pi$ in $B^0\to K^+\pi^- \pi^+\pi^-J/\psi$
decays~\cite{belle_x3872_mass}. In a sample of 657M $B\bar{B}$
pairs a signal of about 90 $X(3872)\to \pi^+\pi^- J/\psi$ events
was observed. Unlike the $B^0\to K^+\pi^- +{charmonium}$ where
$K^+\pi^-$ is mainly from $K^*(892)$ decays, it is evident that
most of the $K\pi$ pairs have a phase space-like distribution,
with little or no signal for $K^*(892)\to K\pi$. Belle reports a
$K^*(892)$ to $K\pi$ non-resonant ratio of
\[
\frac{{\cal B}(B\to (K^+\pi^-)_{K^*(892)}J/\psi)}{{\cal B}(B\to
(K^+\pi^-)_{NR}J/\psi)}<0.55,
\]
at the 90\% C.L. This is another indication that the $X(3872)$
state is not a conventional charmonium state. However, there is no
solid calculation of this above ratio assuming different nature of
the $X(3872)$ state.

BaBar set an upper limit of the $X(3872)$ production rate in the
$B$-meson decays by measuring the momentum distribution of the
inclusive kaon from $B$-meson decays~\cite{PRL96_babar}:
\[
{\cal B}(B^-\to K^- X(3872))<3.2\times 10^{-4}
\]
at the 90\% C.L. Together with all the other measurements on the
product branching fractions ${\cal B}(B^-\to K^- X(3872))\cdot
{\cal B}(X(3872)\to exclusive)$, one gets
\[
2.3\%<{\cal B}(X(3872)\to \pi^+\pi^-J/\psi)<6.6\%,
\]
\[
1.4\times 10^{-4}<{\cal B}(B^-\to K^- X(3872))<3.2\times 10^{-4},
\]
at the 90\% C.L. We find that the decay width of the $X(3872)$ to
$\pi^+\pi^-J/\psi$ is larger and the production rate of the
$X(3872)$ is smaller than conventional charmonium states such as
$\eta_c$, $\psi(2S)$, and $\chi_{c1}$~\cite{PDG}.

\section{The $XYZ$ states near 3.94~GeV}

In 2005, Belle reported observations of three states with masses
near 3940~MeV: the $X(3940)$, seen as a $D^*\bar{D}$ mass peak in
exclusive $e^+e^-\rightarrow J/\psi D^*\bar{D}$
annihilations~\cite{belle_x3940}; the $Y(3940)$, seen as an
$\omega J/\psi$ mass peak in the decay $B\rightarrow K\omega
J/\psi$~\cite{belle_y3940}; and the $Z(3930)$, seen as a
$D\bar{D}$ mass peak in $\gamma\gamma\to D\bar{D}$
events~\cite{belle_z3930}. Of these, only the $Z(3930)$ has been
assigned to a $2^3P_2$ $c\bar{c}$ charmonium state, which is
commonly called the $\chi_{c2}^{\prime}$.

The $X(3940)$ is produced in association with a $J/\psi$ in the
$e^+e^-\to J/\psi X(3940)$ annihilation process, which fixes its
$C$-parity as $C=+1$. Furthermore, the only known charmonium
states that are seen to be produced via the process
$e^+e^-\rightarrow J/\psi (c\bar{c})$ have $J=0$, which provides
some circumstantial evidence that the $X(3940)$ has $J=0$. This,
taken together with the fact that the $X(3940)$ was discovered via
its $D^*\bar{D}$ decay channel and is not seen to decay to
$D\bar{D}$ -- a decay channel that is preferred for $0^{++}$ and
forbidden for $0^{-+}$ -- indicates that $J^{PC}=0^{-+}$ is its
most likely quantum number assignment. The unfilled $0^{-+}$ state
with the closest expected mass value is the $3^1S_0$
$\eta_c^{\prime\prime}$, which potential model predictions put at
4043~MeV (or higher)~\cite{barnes_prd72}, well above the
$X(3940)$'s measured mass of $3942\pm 2\pm
6$~MeV~\cite{belle_prl100202001}.

The $Y(3940)$ mass is well above open-charm mass thresholds for
decays to $D\bar{D}$ or $D^*\bar{D}$ finally states, but was
discovered via its decay to the hidden charm $\omega J/\psi$ final
state. This implies an $\omega J/\psi$ partial width that is much
larger than expectations for usual charmonium.

In a recently reported study of $B\to K D^*\bar{D}$ decays, Belle
searched for, but did not find, a signal for $B\to K Y(3940)$;
$Y(3940)\to D^*\bar{D}$~\cite{belle_0810-0358}. The quoted upper
limit on this mode corresponds to a lower limit on the branching
fraction ratio:
\begin{equation}\label{eq:y3940_2_ddstr}
\frac{{\cal B}(Y(3940)\to  \omega J/\psi)} {{\cal B}(Y(3940)\to
D^{*0}\bar{D^0})}>0.75
\end{equation}
at the 90\% C.L. Likewise, Belle searched for evidence for
$X(3940)\to \omega J/\psi$ by searching for $\omega J/\psi$
systems recoiling from a $J/\psi$ in $e^+e^-\to \omega 2J/\psi$
annihilations~\cite{belle_x3940}. Here no signal is seen and an
upper limit
\begin{equation}\label{eq:x3940_2_wj}
\frac{{\cal B}(X(3940)\to \omega J/\psi)} {{\cal B}(X(3940)\to
D^{*0}\bar{D^0})}<0.60
\end{equation}
was established at the 90\% C.L. These limits would be
contradictory if the $X(3940)$ and the $Y(3940)$ were the same
state seen in different production modes. Thus, the best current
evidence indicates that these two states are distinct.

In 2008, BaBar~\cite{babar_y3940} reported a study of $B\to
K\omega J/\psi$ in which the $\omega J/\psi$ invariant mass
distribution shows a near-threshold peaking that is qualitatively
similar to $Y(3940)$ peak previously reported by Belle. However,
the BaBar values for mass and width derived from fitting their
data are both lower than the corresponding values reported by
Belle and are more precise: $M=3914^{+3.8}_{-3.4}\pm 1.6$~MeV
(BaBar) compared to $3943\pm 11\pm 13$~MeV (Belle), and
$\Gamma=33^{+12}_{-8}\pm 0.6$~MeV (BaBar) compared to $87\pm 22\pm
26$~MeV (Belle). Part of the difference might be attributable to
the larger data sample used by BaBar (350~fb$^{-1}$ compared to
Belle's 253~fb$^{-1}$), which enabled them to use smaller $\omega
J/\psi$ mass bins in their analysis.

\begin{figure}[htb]
\centering
\includegraphics[width=8cm]{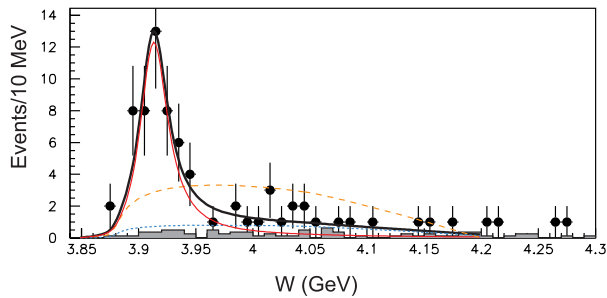}
\caption{The $\omega J/\psi$ mass distribution for selected events
and the fit with a BW function plus a smooth background function
(solid line). The dashed curve shows the fit with no BW term.}
\label{fig:omegajpsi}
\end{figure}

To add more information to the states in this mass region, Belle
observed a dramatic and rather narrow peak, $X(3915)$, in the
cross section for $\gamma\gamma\to \omega
J/\psi$~\cite{belle_y3915} that is consistent with the mass and
width reported for the $Y(3940)$ by the BaBar group. The invariant
mass distribution for the $\omega J/\psi$ candidates produced in
$\gamma\gamma$ collision, shown in Fig.~\ref{fig:omegajpsi}, shows
a sharp peak near threshold and not much else. The statistical
significance of the signal is $7.1\sigma$.

The fit with a BW plus a smooth background function gives results
for the resonance parameters of the $X(3915)$:
\begin{eqnarray}\label{eq:omegajpsfit}
M & = & 3914\pm 4\pm 2~{\rm MeV};\\
\Gamma & = & 28\pm 12^{+2}_{-8}~{\rm MeV}.
\end{eqnarray}

This value for the mass is about $2\sigma$ different from that of
the $Z(3930)$ ($M=3929\pm 5\pm 2$~MeV), indicating that these two
peaks may not be different decay channels of the same state. On
the other hand, there is good agreement between these preliminary
results and the mass and width quoted by BaBar for the $Y(3940)$,
which is also seen in $\omega J/\psi$.

The $X(3915)$ production rate depends on the $J^{P}$ value. Belle
determines
\[
\Gamma_{\gamma\gamma}(X(3915)){\cal B}(X(3915)\to \omega J/\psi)
=69\pm 16^{+7}_{-18}~{\rm eV},
\]
or
\[
\Gamma_{\gamma\gamma}(X(3915)){\cal B}(X(3915)\to \omega J/\psi)
=21\pm 4^{+2}_{-5}~{\rm eV},
\]
for $J^{P}=0^{+}$ or $2^{+}$, respectively.

The nature of the $X(3915)$ is unknown. However, it is very
unlikely to be a charmonium state since the partial width of this
state to $\gamma\gamma$ or $\omega J/\psi$ is too large.

\section{The $Y(4140)$ and $X(4350)$}

Using exclusive $B^+ \to J/\psi \phi K^+$ decays, CDF
Collaboration observed a narrow structure near the $J/\psi \phi$
mass threshold with a statistical significance of
3.8$\sigma$~\cite{CDF}. The mass and width of this structure are
fitted to be $4143.0\pm 2.9(\rm stat)\pm 1.2(\rm syst)~\hbox{MeV}$
and $11.7^{+8.3}_{-5.0}(\rm stat)\pm 3.7(\rm syst)~\hbox{MeV}$
respectively using an $S$-wave relativistic BW function. This new
state, called $Y(4140)$ by the CDF Collaboration, is an isospin
singlet state with positive $C$ and $G$ parities since the quantum
numbers of both $J/\psi$ and $\phi$ are $I^G(J^{PC})=0^-(1^{--})$.
It was argued by the CDF Collaboration that the $Y(4140)$ can not
be a conventional charmonium state, because a charmonium state
with mass about 4143~$\hbox{MeV}$ would dominantly decay into open
charm pairs, and the branching fraction into the double OZI
forbidden modes $J/\psi \phi$ or $J/\psi \omega$ would be
negligible.

There have been a number of different interpretations proposed for
the $Y(4140)$, including a $D_{s}^{\ast+} {D}_{s}^{\ast-}$
molecule~\cite{tanja, liux, ding, namit, liu3, huang, raphael,
molina}, an exotic $1^{-+}$ charmonium hybrid~\cite{namit}, a
$c\bar{c}s\bar{s}$ tetraquark state~\cite{stancu}, or a natural
consequence of the opening of the $\phi J/\psi$
channel~\cite{eef}. There are also arguments that the $Y(4140)$
should not be a conventional charmonium $\chi_{c0}^{\prime\prime}$
or $\chi_{c1}^{\prime\prime}$~\cite{liu2}, nor a scalar
$D_{s}^{\ast+} {D}_{s}^{\ast-}$ molecule~\cite{wangzg, wangzg2}.

The Belle Collaboration searched for this state using the same
process with $772\times 10^6$ $B \bar{B}$ pairs. No significant
signal was found, and the upper limit on the production rate
${\cal B}(B^+\to Y(4140)K^+, Y(4140)\to J/\psi \phi)$ is measured
to be $6\times 10^{-6}$ at the 90\% C.L. Although this upper limit
is lower than the central value of the CDF measurement $(9.0\pm
3.4\pm 2.9)\times 10^{-6}$, it does not contradict with the CDF
measurement considering the large error~\cite{CDF}.

Assuming the $Y(4140)$ is a $D_{s}^{\ast+} {D}_{s}^{\ast-}$
molecule with quantum number $J^{PC}=0^{++}$ or $2^{++}$, the
authors of Ref.~\cite{tanja} predicted a two-photon partial width
of the $Y(4140)$ of the order of 1~keV, which is large and can be
tested with experimental data. The Belle Collaboration searched
for this state in two-photon process~\cite{x4350} to test this
model. This analysis is based on a 825~fb$^{-1}$ data sample
collected at the $\Upsilon(nS)~(n=1,3,4,5)$ resonances. No
$Y(4140)$ signal is observed, and the upper limit on the product
of the two-photon decay width and branching fraction of $Y(4140)
\to \phi J/\psi$ is measured to be $\Gamma_{\gamma
\gamma}(Y(4140)) {\cal B}(Y(4140)\to\phi J/\psi)<39~\hbox{eV}$ for
$J^P=0^+$, or $<5.7~\hbox{eV}$ for $J^P=2^+$ at the 90\% C.L. for
the first time. The upper limit on $\Gamma_{\gamma
\gamma}(Y(4140)) {\cal B}(Y(4140)\to\phi J/\psi)$ from this
experiment is lower than the prediction of $176^{+137}_{-93}$~eV
for $J^{PC}=0^{++}$, $189^{+147}_{-100}$~eV for $J^{PC}=2^{++}$
(calculated by us using the numbers in Ref.~\cite{tanja} and total
width of the $Y(4140)$ from CDF~\cite{CDF}). This disfavors the
scenario of the $Y(4140)$ being a $D_{s}^{\ast+} {D}_{s}^{\ast-}$
molecule with $J^{PC}=0^{++}$ or $2^{++}$.

\begin{figure}[htbp]
  \begin{center}
    \includegraphics[width=8cm]{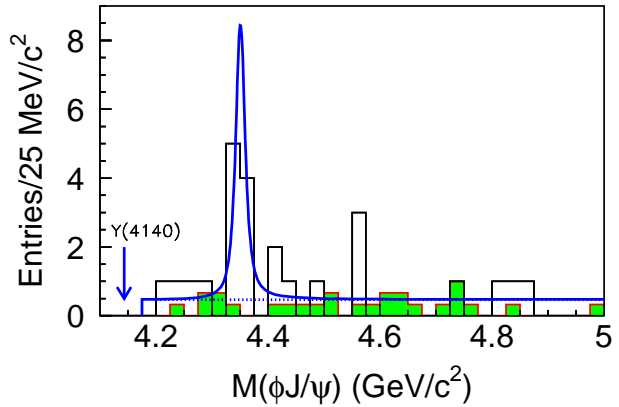}
  \end{center}
\caption{Fit to the $\phi J/\psi$ invariant mass distribution. The
solid line is the best fit, the dashed line is the background, and
the shaded histogram is from normalized $\phi$ and $J/\psi$ mass
sidebands. The arrow shows the position of the $Y(4140)$.}
\label{mkkjpsi-fit2}
\end{figure}

Evidence is reported for a narrow structure at
$4.35~\hbox{GeV}/c^2$ in the $\phi J/\psi$ mass spectrum in the
above two-photon process $\gamma \gamma \to \phi J/\psi$ (see
Fig.~\ref{mkkjpsi-fit2}) in Belle experiment. A signal of
$8.8^{+4.2}_{-3.2}$ events, with statistical significance of
greater than 3.2 standard deviations, is observed. The mass and
natural width of the structure (named as $X(4350)$) are measured
to be $4350.6^{+4.6}_{-5.1}(\rm{stat})\pm
0.7(\rm{syst})~\hbox{MeV}$ and $13.3^{+17.9}_{-9.1}(\rm{stat})\pm
4.1(\rm{syst})~\hbox{MeV}$, respectively. The products of its
two-photon decay width and branching fraction to $\phi J/\psi$ is
measured to be $\Gamma_{\gamma \gamma}(X(4350)) B(X(4350)\to\phi
J/\psi)=6.4^{+3.1}_{-2.3}\pm 1.1~\hbox{eV}$ for $J^P=0^+$, or
$1.5^{+0.7}_{-0.5}\pm 0.3~\hbox{eV}$ for $J^P=2^+$. It is noted
that the mass of this structure is well consistent with the
predicted values of a $c\bar{c}s\bar{s}$ tetraquark state with
$J^{PC}=2^{++}$ in Ref.~\cite{stancu} and a $D^{\ast+}_s
{D}^{\ast-}_{s0}$ molecular state in Ref.~\cite{zhangjr}.

\section{The $Y$ states in $ISR$ processes}

The study of charmonium states via initial state radiation ($ISR$)
at the $B$-factories has proven to be very fruitful. In the
process $e^+e^- \to \gamma_{ISR} \pi^+\pi^-J/\psi$, the BaBar
Collaboration observed the $Y(4260)$~\cite{babary}. This structure
was also observed by the CLEO~\cite{cleoy} and Belle
Collaborations~\cite{belley} with the same technique; moreover,
there is a broad structure near 4.008~GeV in the Belle data. In a
subsequent search for the $Y(4260)$ in the $e^+e^- \to
\gamma_{ISR} \pi^+\pi^-\psi(2S)$ process, BaBar found a structure
at around 4.32~GeV~\cite{babar_pppsp}, while the Belle
Collaboration observed two resonant structures at 4.36~GeV and
4.66~GeV~\cite{belle_pppsp}. Recently, CLEO collected
13.2~pb$^{-1}$ of data at $\sqrt{s}=4.26$~GeV and investigated 16
decay modes with charmonium or light hadrons~\cite{cleoy4260}. The
large $e^+e^- \to \pi^+\pi^-J/\psi$ cross section at this energy
is confirmed.

\begin{figure}[htb]
  \begin{center}
    \includegraphics[width=8cm]{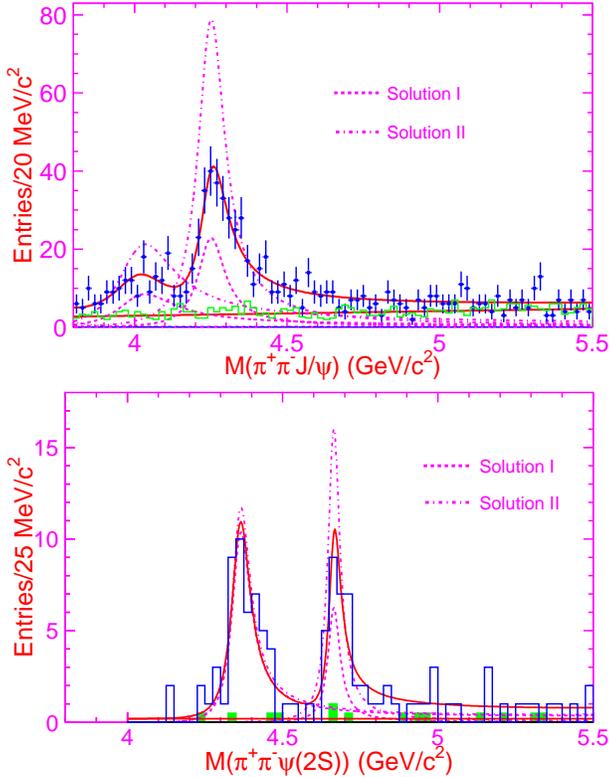}
  \end{center}
  \vspace{-1pc}
\caption{The $\pi^+\pi^-J/\psi$ (upper) and $\pi^+\pi^-\psi(2S)$
(lower) invariant mass distributions and the best fit with two
coherent resonances together with a background term. The data are
from Belle.} \label{mass}
\end{figure}

Figure~\ref{mass} shows the invariant mass distributions of
$\pi^+\pi^-J/\psi$ and $\pi^+\pi^-\psi(2S)$ after all the
selection in Belle data~\cite{belley,belle_pppsp}, together with a
fit with coherent resonance terms and a non-coherent background
term. Table~\ref{tab1} shows the fit results, including the
$Y(4008)$ and $Y(4260)$ from the $\pi^+\pi^-J/\psi$ mode, and the
$Y(4360)$ and $Y(4660)$ from the $\pi^+\pi^-\psi(2S)$ mode. It
should be noted that there are always two solutions in the fit to
each mode, with same mass and width for the resonances but with
very different coupling to $e^+e^-$ pair ($\Gamma_{e^+e^-}$).

\begin{table}[hbt]
\caption{\label{tab1} Fit results of the $\pi^+\pi^-J/\psi$ and
$\pi^+\pi^-\psi(2S)$ invariant mass spectra. The first errors are
statistical and the second systematic. $M$, $\Gamma_{\rm tot}$,
and ${\cal B}\cdot \Gamma_{e^+e^-}$ are the mass (in MeV), total
width (in MeV), product of the branching fraction to hadronic mode
and the $e^+e^-$ partial width (in eV), respectively. $\phi$ is
the relative phase between the two resonances (in degrees).}
\renewcommand{\arraystretch}{1.1}
\begin{center}
{\small
\begin{tabular}{ccc}
\hline\hline
  Parameters & Solution I & Solution II \\
  \hline
  $M(Y(4008))$            & \multicolumn{2}{c}{$4008\pm 40^{+114}_{-28}$}  \\
  $\Gamma_{\rm tot}(Y(4008))$   & \multicolumn{2}{c}{$ 226\pm 44\pm 87$}  \\
  ${\cal B}\cdot \Gamma_{e^+e^-}(Y(4008))$
                  & $5.0\pm 1.4^{+6.1}_{-0.9}$ & $12.4\pm 2.4^{+14.8}_{-1.1}$  \\
  $M(Y(4260))$            & \multicolumn{2}{c}{$4247\pm 12^{+17}_{-32}$} \\
  $\Gamma_{\rm tot}(Y(4260))$   & \multicolumn{2}{c}{$ 108\pm 19\pm 10$} \\
  ${\cal B}\cdot \Gamma_{e^+e^-}(Y(4260))$
                  & $6.0\pm 1.2^{+4.7}_{-0.5}$ & $20.6\pm 2.3^{+9.1}_{-1.7}$ \\
  $\phi$          & $12\pm 29^{+7}_{-98}$ & $-111\pm 7^{+28}_{-31}$
  \\\hline
    $M(Y(4360))$            & \multicolumn{2}{c}{$4361\pm 9\pm 9$} \\
  $\Gamma_{\rm tot}(Y(4360))$   & \multicolumn{2}{c}{$74\pm 15\pm 10$} \\
  ${\cal B}\cdot \Gamma_{e^+e^-}(Y(4360))$
                  & $10.4\pm 1.7\pm 1.5$ & $11.8\pm 1.8\pm 1.4$ \\
  $M(Y(4660))$            & \multicolumn{2}{c}{$4664\pm 11\pm 5$} \\
  $\Gamma_{\rm tot}(Y(4660))$   & \multicolumn{2}{c}{$48\pm 15\pm 3$} \\
  ${\cal B}\cdot \Gamma_{e^+e^-}(Y(4660))$
                  & $3.0\pm 0.9\pm 0.3$ & $7.6\pm 1.8\pm 0.8$ \\
  $\phi$           & $39\pm 30\pm 22$ & $-79\pm 17\pm 20$ \\
  \hline\hline
\end{tabular}}
\end{center}
\end{table}

\begin{figure}[htb]
\centering
\includegraphics[width=7cm]{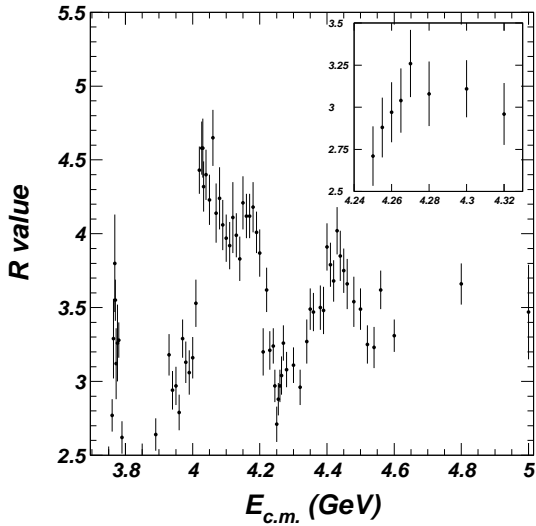}
\caption{The cross section for $e^+e^-\to $~hadrons in the
charmonium region measured by BES (from Ref.~\cite{moxh}).}
\label{fig:bes-Rhad}
\end{figure}

There is only one unassigned $1^{--}$ charmonium state in this
mass region, the $3^3D_1$ level.  This might accommodate the
$Y(4660)$, but there is no room in the spectrum for all the peaks
observed above. A tantalizing feature of all these states is the
absence of corresponding peaking features in the total cross
section for $e^+e^-$ annihilation into hadrons at the same energy
(except the $Y(4008)$ which is close to the $\psi(4040)$).
Figure~\ref{fig:bes-Rhad} shows BES measurements of
$R_{had}=\sigma(e^+e^-\to {\rm hadrons})/\sigma_{QED}(e^+e^-\to
\mu^+\mu^-)$ in the same energy region, where the cross section
exhibits dips near the locations of the $Y(4260)$ and
$Y(4360)$~\cite{bes_Rhad}. (The BES $R_{had}$ measurements do not
span the $Y(4660)$ region.)

The absence of any evidence for the $Y(4260)$ ($Y(4360)$) decays
to open charm implies that the $\pi^+\pi^- J/\psi$ ($\pi^+\pi^-
\psi(2S)$) partial width is large: the analysis of
Ref.~\cite{moxh} gives a 90\% C.L. lower limit $\Gamma(Y(4260)\to
\pi^+\pi^- J/\psi)>508$~keV, which should be compared to the
corresponding $\pi^+\pi^- J/\psi$ partial widths of established
$1^{--}$ charmonium states: 89.1~keV for the $\psi(2S)$ and
44.6~keV for the $\psi(3770)$~\cite{PDG}.

Belle and BaBar have exploited $ISR$ to make measurements of cross
sections for exclusive open-charm final states in this energy
range~\cite{pakhlova_ddstr,pakhlova_dd,pakhlova_lclcbar,pakhlova_ddbarpi,pakhlova_dstrdbarpi}.
The exclusive channels that have been measured so far nearly
saturates the total inclusive cross section, but there is no
evidence for peaking near the masses of the $Y$ states. The one
exception is $e^+e^-\to \Lambda^+_c \Lambda_c^-$, which has a
threshold peak in the vicinity of the $Y(4660)$ peak
mass~\cite{pakhlova_lclcbar}.

The most commonly invoked theoretical explanation for the
$ISR$-produced $1^{--}$ $Y$ states is that they are
$c\bar{c}$-gluon hybrids~\cite{hybrids}, and the relevant
open-charm thresholds for $c\bar{c}$-gluon hybrids are $M_{D^{**}}
+ M_{D}$, where $D^{**}$ designates the low-lying $P$-wave charmed
mesons: $J^P = 0^+$ $D_0(2400)$ and $J^P = 1^+$ $D_1(2420)$. The
prominent decay modes of the $D_0(2400)$ and $D_1(2420)$ are
$D\pi$ and $D^*\pi$, respectively. Therefore, searches for the $Y$
states in both the exclusive $e^+e^-\to D\bar{D}\pi$ and
$D^*\bar{D}\pi$ channels are especially important.

\begin{figure}[htb]
\centering
\includegraphics[width=8cm]{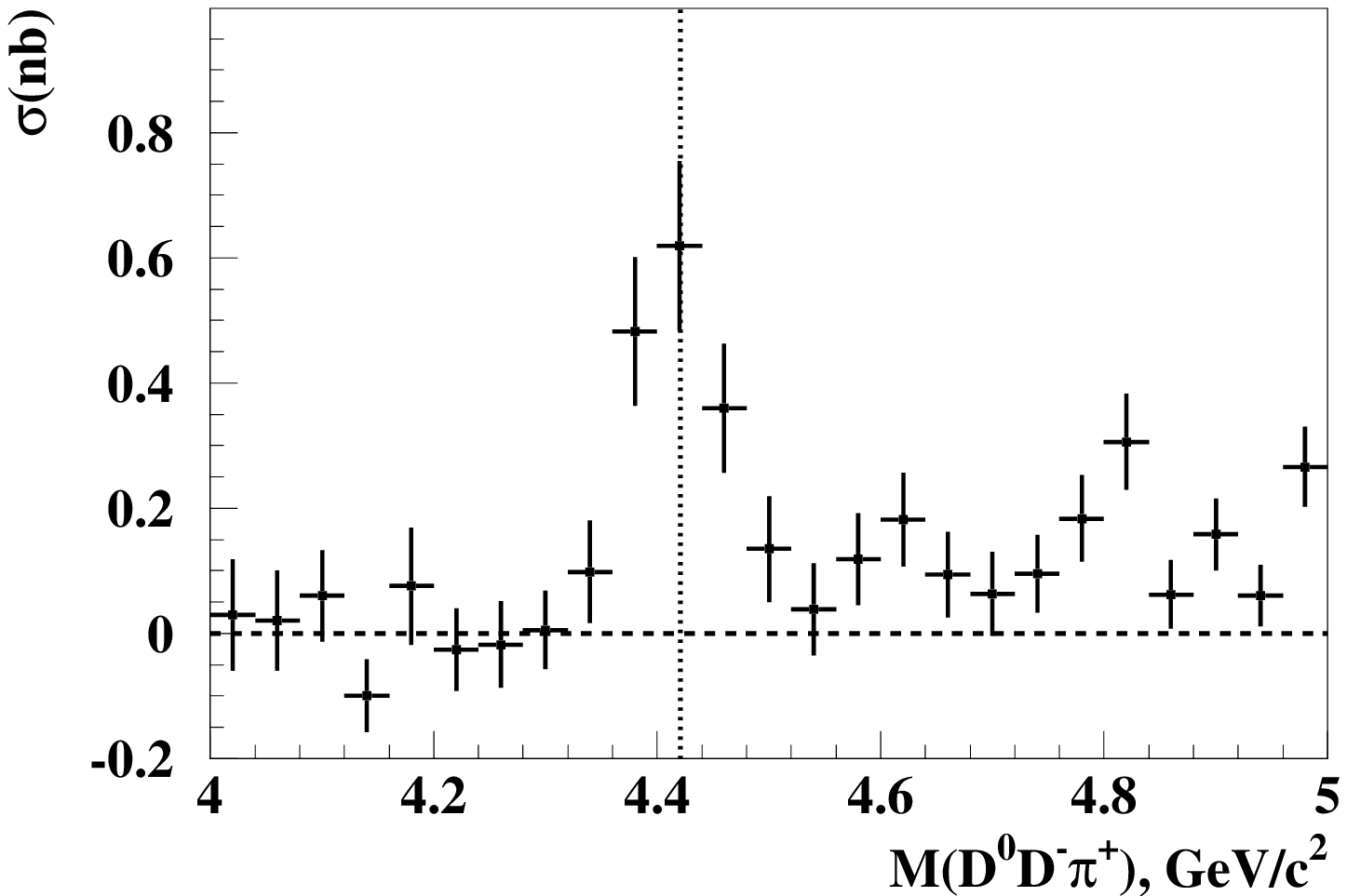}
\caption{ $\sigma(e^+e^- \to  D^0 D^-\pi^+)$ from
Ref.~\cite{pakhlova_ddbarpi}.} \label{fig:y4260_ddbarpi}
\end{figure}

\begin{figure}[htb]
\centering
\includegraphics[width=8cm]{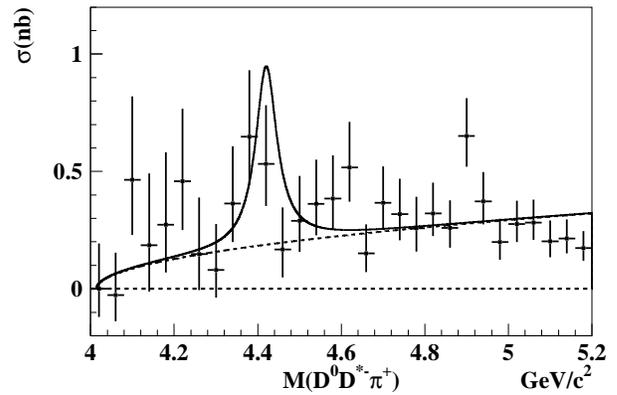}
\caption{ $\sigma(e^+e^- \to  D^*\bar{D}\pi)$ distribution from
Ref.~\cite{pakhlova_dstrdbarpi}. The curve shows the upper limit
on $\psi(4415)$ taking account systematic uncertainties.}
\label{fig:y4260_ddstrpi}
\end{figure}

$\sigma(e^+e^-\to  D^0 D^-\pi)$ and $\sigma(e^+e^-\to
D^{*-}\bar{D}^0 \pi^+)$ measured by Belle are shown in
Figs.~\ref{fig:y4260_ddbarpi} and~\ref{fig:y4260_ddstrpi} The
$\psi(4415)$ signal is strong in $D^0 D^-\pi$ mode while weak in
$D^{*-}\bar{D}^0 \pi^+$ mode. However, the data show no indication
of any of the $Y$ states. A fit to $\sigma(e^+e^-\to
D^{*-}\bar{D}^0 \pi^+)$ using two incoherent Breit-Wigner
functions, one to represent the $Y(4260)$ and the other for the
$\psi(4415)$, plus an incoherent smooth background term give a
90\% C.L. upper limit on ${\cal{B}}(Y(4260)\to  D^0 D^{*-}\pi^+)/
{\cal{B}}(Y(4260)\to \pi^+\pi^- J/\psi)<9$. Similar limits are
obtained for the $Y(4360)$ and $Y(4660)$.

\section{The charged $Z$ states}

Belle's $Z(4430)^+$ signal is the sharp peak in the
$\pi^+\psi(2S)$ invariant mass distribution from $B\to
K\pi^+\psi(2S)$ decays~\cite{belle_z4430}. A fit using a BW
function gives $M=4433\pm 4\pm 2$~MeV and $\Gamma =
45^{+18+30}_{-13-13}$~MeV, with an estimated statistical
significance of more than $6\sigma$. Consistent signals are seen
in various subsets of the data: {\it i.e.} for both the
$\psi(2S)\to \ell^+\ell^-$ and $\psi(2S)\to \pi^+\pi^- J/\psi$
subsamples, the $\psi(2S)$($J/\psi$)$\to e^+e^-$ and $\mu^+\mu^-$
subsamples, etc.

However the BaBar group did not confirm the $Z(4430)^+\rightarrow
\pi^{+}\psi(2S)$ mass peak in their partial wave analysis of $B\to
K\pi\psi(2S)$ decays~\cite{babar_z4430} although statistically
BaBar result does not contradict with Belle's observation. Belle
performed a reanalysis of their data with a similar partial wave
analysis. Specifically, they modelled the $B\to K\pi\psi(2S)$
process as the sum of two-body decays $B\to K_i^*\psi(2S)$, where
$K_i^*$ denotes all of the known $K^*\to K\pi$ resonances that are
kinematically accessible, and both with and without a $B\to K Z$
component, where $Z$ denotes a resonance that decays to
$\pi\psi(2S)$~\cite{belle_z4430_dalitz}.

\begin{figure}[htb]
\centering
\includegraphics[width=7cm]{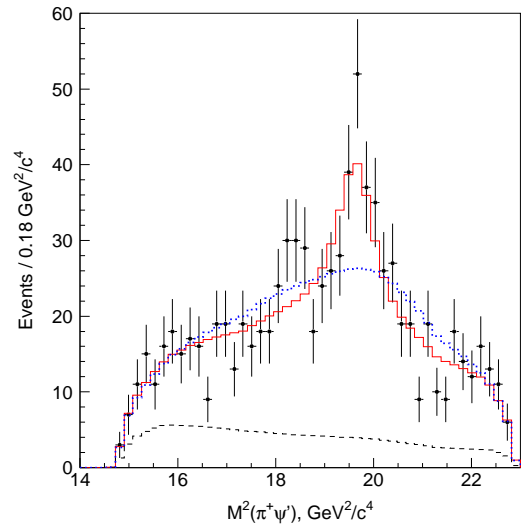}
\caption{The data points show the $M^2(\pi\psi(2S))$ projection of
the Dalitz plot with the $K^*$ bands removed.  The histograms show
the corresponding projections of the fits with and without a $Z\to
\pi\psi(2S)$ resonance term.} \label{fig:z4430_dalitz-analysis}
\end{figure}

The data points in Fig.~\ref{fig:z4430_dalitz-analysis} shows the
$M^2(\pi\psi(2S))$ Dalitz plot projection with the prominent $K^*$
bands removed compared with the results of the fit with no $Z$
resonance, shown as a dashed histogram, and that with a $Z$
resonance, shown as the solid histogram. The fit with the $Z$ is
favored over the fit with no $Z$ by $6.4\sigma$.  The fitted mass,
$M=4443^{+15+19}_{-12-13}$~MeV, agrees within the systematic
errors with the earlier Belle result; the fitted width, $\Gamma =
107^{+86+74}_{-43-56}$~MeV, is larger, but also within the
systematic errors of the previous result. The product branching
fraction from the Dalitz fit: ${\cal B}(B^0\to  K Z^+)\cdot {\cal
B}(Z^+ \to \pi^+\psi(2S)) = (3.2^{+1.8+9.6}_{-0.9-1.6})\times
10^{-5}$ is not in strong contradiction with the BaBar 95\% C.L.
upper limit of $3.1\times 10^{-5}$.

\begin{figure}[htb]
\centering
\includegraphics[width=7cm]{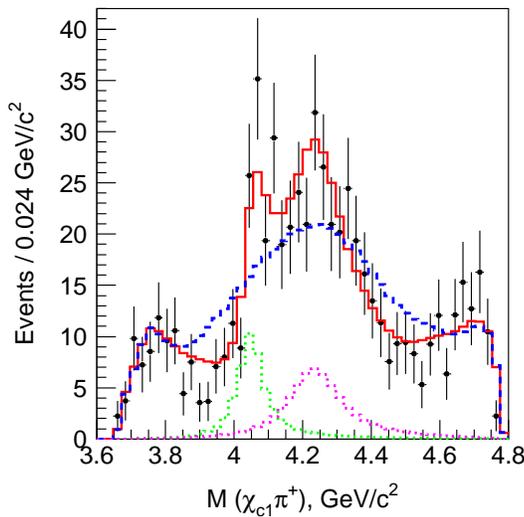}
\caption{The data points show the $M(\pi\chi_{c1})$ projection of
the Dalitz plot with the $K^*$ bands removed. The histograms show
the corresponding projections of the fits with and without the two
$Z\to \pi\chi_{c1}$ resonance terms.}
\label{fig:z4050_dalitz-analysis}
\end{figure}

In addition to the $Z(4430)^+$, Belle has presented results of an
analysis of $B\to K\pi^+\chi_{c1}$ decays that require two
resonant states in the $\pi^+\chi_{c1}$
channel~\cite{belle_z14050}. In this case the kinematically
allowed mass range for the $K\pi$ system extends beyond the
$K^*_3(1780)$ $F$-wave resonance and $S$-, $P$-, $D$- and $F$-wave
terms for the $K\pi$ system are included in the model. The fit
with a single resonance in the $Z\to  \pi\chi_{c1}$ channel is
favored over a fit with only $K^*$ resonances and no $Z$ by more
than $10\sigma$.  Moreover, a fit with two resonances in the
$\pi\chi_{c1}$ channel is favored over the fit with only one $Z$
resonance by $5.7\sigma$. The fitted masses and widths of these
two resonances are: $M_1=4051\pm 14^{+20}_{-41}$~MeV and $\Gamma_1
= 82^{+21+47}_{-17-22}$~MeV and $M_2=4248^{+44+180}_{-29-35}$~MeV
and $\Gamma_2 = 177^{+54+316}_{-39-61}$~MeV. The product branching
fractions have central values similar to that for the $Z(4430)$
but with large errors. Figure~\ref{fig:z4050_dalitz-analysis}
shows the $M(\pi\chi_{c1})$ projection of the Dalitz plot with the
$K^*$ bands excluded and the results of the fit with no $Z\to
\pi\chi_{c1}$ resonances and with two $Z\to \pi\chi_{c1}$
resonances.

Since the $Z$ states have hidden charm and light quarks to make
them decay to charmonium rich final states and with non-zero
charge, if any one of them is confirmed, it is an unambiguous
evidence for state with more than three quarks.

\section{Summary} 

In summary, there are lots of charmonium-like states observed
recently in charmonium mass region but many of them show
properties different from the naive expectation of conventional
charmonium states. All these may suggest the long searching exotic
states have been observed. However, due to limited statistics, the
experimental information on the properties of any of these states
is not enough for us to draw solid conclusion, let alone our poor
knowledge on the QCD prediction of the properties of the exotic
states or the usual charmonium states.

In the near future, BESIII experiment~\cite{bes3} may accumulate
data for center of mass energy between 3 and 4.6~GeV, this will
contribute to the understanding of some of these states discussed
above; the Belle II experiment~\cite{belle2} under construction,
with about 50~ab$^{-1}$ data accumulated, will surely improve our
understanding of all these states.

\section*{Acknowledgment}

We borrowed many material from arXiv:0909.2713 by Steve Olsen,
where the same topic was discussed based on the same experimental
information. We acknowledge Dr.~M.~Nielsen for helpful
discussions, and thank the organizers for their kind invitation
and congratulate them for a successful workshop. This work is
supported in part by National Natural Science Foundation of China
under Contract Nos. 10825524 and 10935008.



\end{document}